\documentclass[journal, twocolumn, 10pt, letterpaper]{IEEEtran}
\usepackage{graphics}
\usepackage{amsmath}
\usepackage{cite}
\usepackage{amsthm}
\usepackage{amssymb}
\usepackage{color}
\usepackage{trig}
\usepackage{tikz}

\newtheorem{definition}{Definition} [section]
\newtheorem{theorem}{Theorem}[section]
\newtheorem{lemma}{lemma}[section]

\newtheorem{example}{Example}[section]

\begin{document}

\title{Electricity Pooling Markets with Strategic Producers Possessing Asymmetric Information II: Inelastic Demand}
\author{%
Mohammad Rasouli and Demosthenis Teneketzis, Fellow IEEE \\
Email: rasouli@umich.edu, teneket@umich.edu\\
Department of EECS\\
University of Michigan\\
Ann Arbor, MI , 48109-2122
}
\date{}
\maketitle

\begin{abstract}
\label{abstract}
In the restructured electricity industry, electricity pooling markets are an oligopoly with strategic producers possessing private information (private production cost function). We focus on pooling markets where aggregate demand is represented by a non-strategic agent.

Inelasticity of demand is a main difficulty in electricity markets which can potentially result in market failure and high prices. We consider demand to be inelastic.

We propose a market mechanism that has the following features. (F1) It is individually rational. (F2) It is budget balanced. (F3) It is price efficient, that is, at equilibrium the price of electricity is equal to the marginal cost of production.  (F4) The energy production profile corresponding to every non-zero Nash equilibrium of the game induced by the mechanism is a solution of the corresponding centralized problem where the objective is the maximization of the sum of the producers' and consumers' utilities.

We identify some open problems associated with our approach to electricity pooling markets.
\end{abstract}


\section{Introduction}
\label{Introduction}

\subsection{Motivation}

Electricity markets are the heart of the industry restructuring. Electricity as a trading commodity has unique features \cite{Wilson} due to the physical laws (KVL, KCL) governing the power flow, the lines' capacity constraints, and the fact that electric energy cannot be stored. Pooling markets have been proposed as a solution to energy trading.

In a previous paper \cite{rasouli1}, we presented the state of the art in the design of electricity pooling markets and discussed three existing approaches to electricity pooling markets, namely, the Cournot approach \cite{Varaiya1}, \cite{Motto}, \cite{YaoOren},\cite{Ventosa}, \cite{SioshansiOren}, the Bertrand approach \cite{Ventosa} and the supply function approach \cite{SioshansiOren}, \cite{Contreras}, \cite{Borenstein2}, \cite{Ott}, \cite{Faruqui}. In \cite{rasouli1}, we proposed a new approach to electricity pooling markets with strategic producers possessing asymmetric information, non-strategic consumers, and elastic demand.

In this paper, we study electricity pooling markets with strategic producers possessing asymmetric information, non-strategic consumers, and inelastic demand. Inelastic demand is an appropriate assumption for real-time markets as well as congested day-ahead markets \cite{Bushnell, Kirschen}.

Electricity pooling markets with inelastic demand result in an important and challenging problem in the restructuring of electricity markets. Reference \cite{Borenstein1} argues that the fundamental problem with electricity markets is that the demand is almost completely insensitive to price fluctuations, while supply faces binding constraints at peak times, and storage is prohibitively costly. As a result, in a market with no demand elasticity and strict production constraints, a firm with even a small percentage of the market could exercise extreme market power when demand is high. For example, in a case that ISO needs 97 percent of all generators running to meet demand, a firm that owns 6 percent of capacity can exercise a great deal of market power. Reference \cite{Borenstein2} studies and measures this market power. Inelasticity of demand was the main reason for market failure in electricity markets e.g. the California Independent System Operator/ Power Exchange (ISO/PX) experience \cite{Borenstein1}. 


As a solution to problems associated wth inelastic demand, \cite{Borenstein1} proposes long-term contracting combined with retail market to introduce elasticity to the electricity market.

In this paper, we consider pooling markets with inelastic demand and present a mechanism for production allocation among strategic producers that has the following properties. (1) It induces price-taking behavior among the producers (i.e. it deals with the market power even small strategic producers could exercise) (2) It incentivizes strategic producers to collectively meet the demand. Price taking behavior is achieved by paying each producer a price that is determined by the price proposals of the other producers (i.e. the price a producer is paid is independent of its own price proposal). Furthermore, to limit the differences among the price proposals of producers, we consider penalty terms (taxes) for differences in price proposals. In the case of inelastic demand, total production must not exceed the demand because the ISO wants to maximize the social welfare (defined in Section \ref{Model}). The fixed utility of demand is part of the social welfare, and the ISO does not want the demand to pay for overproduction (i.e. energy that is not used). To induce total energy production equal to the fixed demand, we consider penalty terms for underproduction and overproduction.

The key difference between the mechanisms proposed in \cite{rasouli1} and the mechanism presented in this paper are in the incentives provided to strategic producers so as to participate in the electricity pooling market. These differences are discussed in Section \ref{Interpretation of the Mechanism}.

\subsection{Contributions of the paper}
In this paper we focus on electricity pooling markets with strategic producers possessing private information, non-strategic (price-taking) inelastic demand, and no transmission constraints. We adopt Nash Equilibrium (NE) as a solution/equilibrium concept; the interpretation of NE is the same as in \cite{demos3}- \cite{sharma3}.

We propose a market mechanism which has the following features. \textbf{(F1)} It is individually rational. That is, strategic producers voluntarily participate in the pooling market. \textbf{(F2)} It is budget balanced. That is the mechanism does not create any budget surplus or any budget deficit. \textbf{(F3)} It is price efficient. That is, the demand is paying a price equal to the marginal cost of producing the next one unit of energy. \textbf{(F4)} The energy production profile corresponding to every non-zero NE of the game induced by the mechanism is a solution of the corresponding centralized problem, i.e. the problem the ISO would solve if it had access to the producers' private information. Furthermore, if zero production is the only NE of the game induced by the mechanism, the energy production profile corresponding to that is a solution of the corresponding centralized optimization problem. For discussion on these features see \cite{rasouli1}.

The mechanism/game form presented in this report is distinctly different from currently existing mechanisms for electricity pooling markets. There are two key differences between our mechanism and currently existing mechanisms:
\begin{enumerate}
\item in terms of the type of information exchange (the message space of the mechanism);
\item in terms of the performance.
\end{enumerate}
We elaborated on these key differences in \cite{rasouli1}.

\subsection{Organization}
The rest of the paper is organized as following. The model of the market analyzed/studied in this paper is introduced in Section \ref{Model}. The objective is presented in section \ref{Objective}. The centralized optimization problem associated with the model and the objective is presented in Section \ref{Centralized Problem}. The mechanism for inelastic demand and its analysis are presented in Section \ref{Inelastic}. Discussion of open problems associated with our approach to electricity pooling markets appears in Section \ref{Conclusion}. The proofs of our results are presented in Appendices \ref{appendix_A}-\ref{appendix_B}. An example illustrating our approach and results appears in Appendix \ref{appendix_C}.

\section{The Model}
\label{Model}

We consider a pooling market consisting of an ISO, $N$ producers, and consumers who are represented by their aggregate demand. Let $I=\{1,2,...,N\}$ denote the set of producers. We make the following assumptions:

(A1) The number of of producers, $N$, is fixed and common knowledge among the ISO, producers and consumers; furthermore, $N>3$.

(A2) Producers are strategic and self-profit maximizers.

(A3) Each producer $i$ has a fixed capacity $x_i > 0$, $i=1,2,...,N$, which is common knowledge among the producers, the ISO and the consumers.

(A4) The cost function $C_i(.)$, $i=1, 2, ..., N$, of energy production is producer $i$'s private information.  Also, $C_i(.)  \in \mathcal{C}_i$, where the function space $\mathcal{C}_i$ is common knowledge among producers and the ISO.

(A5) The functions $C_i(.)$, $i=1, 2, ..., N$, are convex; furthermore, for all $i$, $i=1,2, ..., N$, 
\begin{eqnarray}
\label{C_0}
C_i(0)=0,\\
\label {C_first_der}
C_i^{'}(e_i) > 0,\\
\label{C_second_der}
C_i^{''}(e_i) > 0,
\end{eqnarray}
for all $e_i>0$, where $e_i$ denotes the amount of energy produced by producer $i$, and $C_i^{'}(.)$ and $C_i^{''}(.)$ denote the first and second derivatives, respectively, of $C_i(.)$. 

(A6) Producer $i$'s utility function is
\begin{equation} 
\label{utility_consumers}
u_i(e_i, t_i)= -C_i(e_i)+t_i
\end{equation}
where $t_i$ denotes the amount of money producer $i$ receives for the energy it produces. 

(A7) The demand is inelastic. Its amount, $D_0$, is fixed and common knowledge among the ISO, the producers and the consumers. $D_0$ is less than or equal to the sum of the producers' capacities.

(A8) Consumers are non-strategic, their total utility is
\begin{equation}
\label{utility_demand}
u_{D_0}-\sum_{i\in I} t_i
\end{equation}
where $u_{D_0}$ is the fixed utility the consumers gain from consuming $D_0$ amount of energy and is a common knowledge among ISO, producers and consumers. $\sum_{i\in I} t_i$ denotes the amount of money demand should pay to the producers for the energy consumed.

(A9) The ISO is a social welfare maximizer. From (\ref{utility_consumers}) and (\ref{utility_demand}) the social welfare function for the case of inelastic demand is 
\begin{eqnarray}
\label{social_welfare_inelastic}
W_1(e_1, e_2, ..., e_N) &=& u_{D_0}-\sum_{i\in I} t_i-\sum_{i\in I} C_i(e_i)+\sum_{i\in I} t_i \nonumber \\
&=& u_{D_0}-\sum_{i\in I} C_i(e_i). 
\end{eqnarray}
Since $u_{D_0}$ is constant and common knowledge among the producers, the consumers and the ISO, maximizing the social welfare function given by is equivalent to maximizing
\begin{eqnarray}
\begin{aligned}
\label{social_welfare_inelastic_2}
W_2(e_1, e_2, ..., e_N) = -\sum_{i\in I} C_i(e_i). 
\end{aligned}
\end{eqnarray}

(A10) No transmission constraints are taken into account in the energy distribution. 

Assumptions (A1)-(A6) and (A10) appear in \cite{rasouli1} and they are discussed in the same reference. Here we discuss Assumptions (A7)-(A9).

Inelastic demand can not change its amount of consumption ((A7)). The same observation is made in reference \cite{baldick}. Inelasticity of the demand usually happens at peak demand hours and is a more compatible assumption with real-time markets. The amount of inelastic demand, $D_0$, is the aggregate of the demand bids submitted to the pooling market by consumers/retailers. This value of demand is then broadcasted to the producers by the ISO. Assuming $\sum_{i\in I}{e_i} \ge D_0$, (A7), ensures that the producers can meet the demand with their current capacity. This assumption is essential for market clearing.

(A8) implies that the demand is non-strategic. Demand does not bid in the market and does not decide on its amount of production and its payment. Reference \cite{elmaghrabyOren} has the same assumption in the pooling market and argues that it is consistent with most currently operating and proposed power auctions.

Considering the form of the producers' utilities as well as the consumers' utility, the non-profitmaker ISO aims to maximize the social welfare defined in (\ref{social_welfare_inelastic}). Note that for inelastic demand, the excess production does not affect the utility of the consumers and therefore, assuming $\sum_{i\in I} e_i \geq D_0$, maximizing the social welfare is equivalent to minimizing Eq. (\ref{social_welfare_inelastic_2}), the total cost of production.

\section{Objective and Method of Approach}
\label{Objective}
The ISO's objective is to maximize the social welfare function given by Eq. (\ref{social_welfare_inelastic}) under the constraints imposed by (A1)-(A10) along with the requirement that the capacity constraints of the producers are satisfied and total production exceeds inelastic demand $D_0$.

To achieve this objective, we proceed as follows. We first consider the centralized optimization problem the ISO would solve if he had perfect knowledge of the cost functions, $C_i(.)$, $i=1,2,...,N$. The solution of this centralized problem would give the best possible performance the ISO can achieve. Afterwards, we design a mechanism/game form that possesses properties (F1)-(F4). The above properties are obtained via the creation of a tax function which incentivizes each strategic producer to align his own individual objective with the social welfare. The specification and interpretation of the mechanism and its tax function appears in Section \ref{Inelastic}.

\section{The Centralized Problem}
\label{Centralized Problem}

Because of Eq. (\ref{social_welfare_inelastic_2}), the ISO's centralized problem is 
\begin{eqnarray}
\label{MAX2}
\max_{e_i , i\in I} & & -\sum _ {i\in I}  C_i (e_i)\nonumber\\
s.t. & & 0 \leq e_i \leq x_i \nonumber \\
	& & \sum_{i\in I} e_i \geq D_0.
\end{eqnarray}
We call the above problem \textit{\textbf{MAX1}}.

Assumptions (A3), (A5) and (A7) imply that in \textit{\textbf{MAX1}}, the objective function is strictly concave and the set of feasible solutions is non-empty, convex and compact. Therefore, \textit{\textbf{MAX1}} has a unique solution, and the Karush-Kuhn-Tucker (KKT) conditions are necessary and sufficient for optimality. The KKT conditions are useful for the analysis of the mechanism proposed in this paper. That is why they are presented in Appendix \ref{appendix_A}.

\section{The Mechanism for Inelastic Demand}
\label{Inelastic}
We first specify the mechanism, then we interpret its elements, mainly the tax function, and, finally, we study its properties. We illustrate the mechanism via two examples that appear in Appendix \ref{appendix_C}.

\subsection{Specification of the Mechanism}
\label{elastic_mechanism}

A game form/mechanism is described by $(\mathcal{M}, h)$, where $\mathcal{M}$ is the message/strategy space and $h: \mathcal{M} \rightarrow \mathcal{A}$ is the outcome function from the message space to the space $\mathcal{A}$ of allocations.

We consider the following mechanism. 

\textbf{\underline{Message space}} Let $\mathcal{M}$ be
\begin{equation}
\mathcal{M}:=(\mathcal{M}_1 \otimes \mathcal{M}_2 \otimes ... \otimes \mathcal{M}_N), 
\end{equation}
where $\mathcal{M}_i$ is producer $i$'s message space,
\begin{equation}
\mathcal{M}_i:= [0, x_i]\times \mathbb{R}_{+}, i\in I
\end{equation}
and $m_i\in \mathcal{M}_i$ is of the form 
\begin{equation}
m_i=(\hat{e}_i, p_i)
\end{equation}
where $\hat{e}_i$ denotes the amount of energy producer $i$ proposes to produce, and $p_i$ denotes the price producer $i$ proposes to be paid per unit of energy it produces. Note that $\hat{e}_i$ is restricted by $0 \le \hat{e}_i \le x_i$ and $p_i$ is restricted by $p_i \ge 0$.

\textbf{\underline{Allocation Space}}  Let $\mathcal{A}$ be
\begin{equation}
\mathcal{A}:=(\mathcal{A}_1 \otimes \mathcal{A}_2 \otimes ... \otimes \mathcal{A}_N), 
\end{equation}
where $\mathcal{A}_i$ is producer $i$'s allocation space
\begin{equation}
\mathcal{A}_i:= [0, x_i]\times \mathbb{R}, i\in I,
\end{equation}
and $a_i\in \mathcal{A}_i$ is of the form 
\begin{equation}
a_i=(e_i, t_i),
\end{equation}
where ${e}_i$ denotes the amount of energy producer $i$ is scheduled to produce, and $t_i$ denotes the subsidy (respectively tax) producer $i$ should receive (respectively pay).

\underline{\textbf{Outcome function}} $h: \mathcal{M} \rightarrow \mathcal{A}$

For each $\textbf{m}:=(m_1, m_2, ...,m_N) \in \mathcal{M}$ we have 
\begin{equation}
h(\textbf{m})=({\textbf{e}}, \textbf{t})=({e}_1,...,{e}_N,{t}_1,...,{t}_N),
\end{equation}
where
\begin{eqnarray}
\label{mechanism1}
{e_i} &=& \hat{e_i} \\
\label{tax_form}
t_i &=& p_{i+1} e_i - (p_i-p_{i+1})^2 - 2 p_i \zeta^2\\
\zeta &=& |D_0-\sum_{i\in I} e_i| \\
p_{N+1}&:=&p_{1}.
\end{eqnarray}
We proceed to interpret and analyze the properties of the proposed mechanism.

\subsection{Interpretation of the Mechanism}
\label{Interpretation of the Mechanism}
Since the designer of the mechanism, i.e. ISO, can not alter the producers' cost functions, $C_i(.)$, $i=1,2,...,N$, even if he knew their functional form, the only way it can achieve his objective is through the use of appropriate tax incentives/tax functions. The tax incentive of our mechanism for produce $i$ consists of two components, that is, 
\begin{eqnarray}
t_i &=& t_{i,1} + t_{i,2},
\end{eqnarray}
where
\begin{eqnarray}
\label{t_i,1}
t_{i,1}  &=& p_{i+1} e_i, \\
t_{i,2}  &=& -(p_{i}-p_{i+1})^2 - 2 p_{i} \zeta^2.
\end{eqnarray}

The term $t_{i,1}$ specifies the amount producer $i$ receives for its production $e_i$ from the demand side. It is important to note that the price per unit of electricity energy that a producer is paid is determined by the message/proposal of other producers. Thus, a producer does not control the price it is paid per unit of electricity it provides. This term induces price-taking behavior on behalf of the producers.

The term $t_{i,2} $ provides the following incentives to the producers: (1) To bid/propose the same price per unit of produced energy (2) To collectively propose a total electricity supply that is equal to the demand.

The incentive provided to all producers to bid the same price per unit of produced energy is described by the term $-(p_i-p_{i+1})^2$, which is a positive punishment (tax) paid by producer $i$ for deviating from the price proposal of producer $i+1$.

The incentive provided to all producers to collectively propose a total production that is equal to the demand, $D_0$, is captured by the term $-2 p_{i} \zeta^2$ which is a penalty for both underproduction and overproduction.

$t_{i,2}$ can be thought of as tax payments the ISO collects from the producers in order to align their productions with the social welfare maximizing production.


\subsection {Properties of the Mechanism}

The properties possessed by the proposed mechanism are described by Theorems (\ref{trivial_NE_inelastic})-(\ref{budget_efficiency_inelastic}) and lemma (\ref{lemma1_inelastic}). The proof of all these properties are presented in Appendix \ref{appendix_B}.

\begin{theorem}
\label{trivial_NE_inelastic}
(Existence of NE) One of the NE of the game induced by the proposed mechanism is $m_i^* = (0,0)$, for all $i\in I$. The corresponding production profile and taxes at this equilibrium are zero.
\end{theorem}

\begin{definition} We call the NE where for all $i\in I$, $m_i^*=(0, 0)$ a trivial Nash Equilibrium. We call any other NE of the game induced by the proposed mechanism a non-trivial NE.
\end{definition}

\begin{theorem} \label{feasibility_inelastic} (FEASIBILITY) If $\textbf{m}^*= (\hat{\textbf{e}}^*, \textbf{p}^*) = (\hat{e}_1^*, \hat{e}_2^*, ..., \hat{e}_N^*, p_1^*, p_2^*, ... , p_N^*)$ is a non-trivial NE of the game induced by the proposed mechanism, then the corresponding allocation $\textbf{e}^*$ is a feasible solution of problem \textbf{MAX1}, i.e.
\begin{eqnarray}
\label{feasibility_1}
D_0 - \sum_{i\in I} e_i^* = 0,
\end{eqnarray}
or, equivalently,
\begin{eqnarray}
\label{zetaplus0}
\zeta^*=0.
\end{eqnarray}
\end{theorem} 

\begin{lemma}
\label{lemma1_inelastic}
Let $\textbf{m}^*=(\textbf{e}^*, \textbf{p}^*)$ be a non-trivial NE of the game induced by the mechanism. Then for every producer $i \in I$ we have,
\begin{eqnarray}
\label{price_same}
p_i^*=p_{i+1}^*=p^*\\
\label{vareps0}
p^*(\sum_{i\in I}e^*_i-D_0) =0\\
\label{tax_equ}
t_i^*=p^* e_i^*\\
\label{tax_der_equ}
\frac{\partial t_i}{\partial e_i} |_{\textbf{m}^*} = p^*.
\end{eqnarray}
\end{lemma}

\begin{theorem} 
\label{Nash_Implementation_inelastic}
Consider any non-trivial NE $\textbf{m}^*$ of the game induced by the mechanism. Then, the production profile $\textbf{e}^*$ corresponding to $\textbf{m}^*$ is an optimal solution of the centralized problem $\textbf{MAX1}$.
\end{theorem}

\begin{theorem} 
\label{Strong_Implementation_inelastic} Consider a solution $\textbf{e}^*=(e_1^*, e_2^*, ..., e_N^*)$ of the centralized problem $\textbf{\textit{MAX1}}$. There exists a non-trivial NE, $\textbf{m}^*= (\hat{e}_1^*, \hat{e}_2^*, ..., \hat{e}_N^*, p_1^*,p_2^*, ... ,p_N^*)$ of the game induced by the mechanism such that the production profile corresponding to $\textbf{m}^*$ is equal to $\textbf{e}^*$.
\end{theorem}

Theorems (\ref{trivial_NE_inelastic}) and (\ref{Strong_Implementation_inelastic}) show the game induced by the mechanism has a set of NE consisting of all the solutions to the centralized problem plus a trivial NE of all zero prices and all zero productions. Since problem $\textbf{MAX1}$ has exactly one solution, which is of course with non-zero aggregate production, we can infer that every game induced by the mechanism has exactly two NE, one trivial and one non-trivial.

\begin{theorem}
\label{Individual_Rationality_inelastic} (INDIVIDUAL RATIONALITY) The proposed game form is individually rational, that is at every NE of the game induced by the mechanism, the corresponding allocation $(\textbf{e}^*, \textbf{t}^*)$ is weakly preferred by all producers to the initial allocation $(\textbf{0} , \textbf{0})$.
\end{theorem}

Note that by Theorem (\ref{Individual_Rationality_inelastic}), the trivial NE, i.e. zero productions and zero payments, is Pareto dominated by any other NE of the game.

\begin{theorem}
\label{budget_balance_inelastic}
(BUDGET BALANCE) The mechanism is budget balanced both at equilibrium and off equilibrium, that is the payments from all producers and consumers sum up to zero.
\end{theorem}

\begin{theorem} 
\label{budget_efficiency_inelastic}
(PRICE EFFICIENCY) The mechanism is price efficient; that is at equilibrium, the demand is paying a price equal to the marginal cost of the next one unit of production.
\end{theorem}

The result of Theorem (\ref{budget_efficiency_inelastic}) shows that the game induced by the proposed mechanism incentivizes the producers to reveal the true marginal cost of production of the system at equilibrium.
\section{Conclusion and Reflections}
\label{Conclusion}
Electricity restructuring has changed the industry from a monopoly into an oligopoly where energy producers are strategic players with private information and market power. In this new environment, electricity can be traded through bilateral contracts or pooling markets. In this paper we focused on pooling markets with strategic producers possessing private information, non-strategic consumers with inelastic demand, and no transmission constraints. We designed a mechanism which has properties (F1)-(F4). We observed that the ISO's centralized problem of social welfare maximization has a unique solution and this unique solution is the same as the allocation corresponding to the unique non-trivial NE of the game induced by the mechanism. Achieving these features all together distinguishes our mechanism from any other market design available in the literature. It is worth noting that price efficiency is achieved even though customers (represented by their aggregate demand) are not strategic and the demand side is inelastic. 

The intuition behind the proposed mechanism is the following. The primary considerations/goals in the design of a mechanism for the problem considered in this paper are : (1) Dealing with the strategic producers' market power resulting from the inelasticity of the demand and the non-strategic consumers. (2) Incentivizing the strategic producers to collectively meet the demand. The producers' market power is addressed by offering each producer a price per unit of its produced energy that does not depend on its own price proposal, that is, it is not under its own control; such an offer induces price-taking behavior among the producers. Incentivization of the strategic producers to collectively meet the demand is achieved by the specification of taxes that penalize over-production and under-production. In addition to achieving the above goals, the proposed mechanism possesses other desirable features such as (F1)-(F4). 

We now contrast the mechanism in this paper with the market mechanism for elastic demand proposed by the authors in \cite{rasouli1}. The differences appear in the specification of the tax function which provides incentives to strategic producers to align their interests/goals with those of the ISO. There are two key differences in the tax function of the elastic and inelastic demand. (1) In the case of elastic demand, the price consumers pay defines the amount of energy they consume. Thus, each consumer must pay an amount of money that is equal to the amount of energy it consumes times the price per unit of energy. This implies that in order to achieve budget balance, money must be exchanged among the strategic producers. That is, we must have $\sum_{i\in I} t_i= p \sum_{i\in I} e_i$. This requirement necessitates the presence of the term $t_{i,3}=-(p_{i+1}-p_{i+2})^2$ in the tax function of every strategic producer. The term $t_{i,3}$ does not depend on producer $i$'s message $m_{i}=(e_i,p_i)$, thus, it does not alter its strategic behavior, furthermore the presence of the terms $t_{i,3}$, $i\in I$, collectively results in budget balance. In the case of inelastic demand, consumers are willing to pay any amount of money to get the desired amount of energy. Therefore, budget balance can be achieved without any money exchanged among the strategic producers, and the presence of the terms that are the analogue to the $t_{i,3}$, $i\in I$, is in not necessary. The consumers' utility is $u_d(D_0)-\sum_{i\in I}(t_{i,1}+t_{i,2})$, the total subsidy received by the strategic producers is $\sum_{i\in I} (t_{i,1}+t_{i,2})$ and the mechanism is budget balance. (2) The two mechanisms are different in the way agents contributing to the tax payment imposed for feasible allocation at equilibrium. In both cases, elastic and inelastic demand, the total tax $\sum_{i\in I} p_i \zeta^2$ is imposed on all strategic producers to collectively propose production profiles which are feasible at equilibrium. However, the contribution of each producer to the tax $\sum_{i\in I} p_i\zeta^2$ is different in the two cases. In the case of elastic demand, producer $i$'s contribution is $p_{i+1}\zeta^2$; when demand is inelastic, producer $i$'s contribution is $p_{i}\zeta^2$. The reason for this difference is the following. In order for the tax $\sum_{i\in I} p_i\zeta^2$ to achieve its goal, each producer's contribution must depend on its proposed price, so that its contribution depends on its whole message proposed i.e. production and price. In the case of elastic demand, $\zeta=|D(\overline{p})-\sum_{i\in I}e_i|$ does depend on the proposed price vector $\textbf{p}:=(p_1,p_2,...,p_N)$, hence on $p_i$, and the goal of the tax $\sum_{i\in I}\zeta^2$ is achieved by defining producer $i$'s contribution to this tax to be $p_{i+1}\zeta^2$. \footnote{In this case, designing producer $i$'s contribution to be $p_i\zeta^2$ does not result in feasible allocation at equilibrium. Thus the tax term $\sum_{i\in I} p_i\zeta^2$ imposed on the strategic producers does not incentivize them to collectively propose feasible allocation at equilibrium.} In the case of inelastic demand, $\zeta=|D_0-\sum_{i\in I}e_i|$ does not depend on prices and hence $p_i$, thus, if  $\sum_{i\in I} p_i \zeta^2$ is the total tax required for achieving feasible production profiles at equilibrium, producer $i$'s contribution must be $p_i \zeta^2$.

The game form presented in this paper ensures that the desired allocations are achieved at equilibria without specifying how an equilibrium is reached. In \cite{rasouli1}, we discussed the difficulties associated with the discovery of algorithms (tatonment processes) that guarantee convergence to NE of the game induced by the mechanism.
 
Future problems along this line of research presented in this paper include the incorporation of transmission constraints into the model, the design of dynamic mechanisms for markets that evolve over time, and the design of market mechanisms that work under both elastic and inelastic demand.

\textbf{Acknowledgment}: This research was supported in part by NSF Grant CNS-1238962. The authors thank Hamidreza Tavafoghi for many useful discussions.

\begin{appendices}
\section{\textbf{The Karush-Kuhn-Tacker condition for problem \textit{MAX1}}}
\label{appendix_A}
The Lagrangian for \textit{\textbf{MAX1}} is
\begin{eqnarray}
\label{Lagrangian_MAX2}
\mathcal{L}_{MAX1} &=& - \sum_{i\in I} C_i(e_i) +\lambda(\sum_{i\in I} e_i-D_0)\nonumber\\
&&+\sum_{i\in I}\mu_i (x_i-e_i)+\sum_{i\in I}\nu_i e_i,
\end{eqnarray}
and KKT conditions are
\begin{eqnarray}
\label{KKT}
\label{KKT_MAX1_first}
\left. \frac{\partial C_i}{\partial e_i} \right|_{e^*} - \lambda^* + \mu_i^*-\nu_i^* &=& 0 \qquad \forall i\in I,\\ 
\lambda^*(\sum_i e_i^*-D_0)&=&0,\\
{\mu}^*_i(x_i-e_i^*)&=&0\qquad \forall i\in I,\\
{\nu}^*_i e_i^* &=&0\qquad \forall i\in I,\\
{\lambda}_i^* &\geq& 0,\\
{\nu}_i^* &\geq& 0\qquad \forall i\in I,\\
\label{KKT_MAX1_last}
{\mu}_i^* &\geq&0\qquad \forall i\in I.
\end{eqnarray}

\section{\textbf{Proof of Theorems (\ref{trivial_NE_inelastic})-(\ref{budget_efficiency_inelastic}) and of lemma (\ref{lemma1_inelastic})}}
\label{appendix_B}

\textbf{Proof of Theorem \ref{trivial_NE_inelastic}} 
Consider producer $i$ and let $m_{j}^* = (0,0) \quad \forall j\neq i$. Now note that since $p_{i+1}^*=0$, producer $i$ is paid no compensation for his production, i.e. $p_{i+1}^* e_i$=0. Other terms in the utility function including the cost of production and penalty terms in the tax function are non-positive. Therefore, the utility of producer $i$ is always non-positive and proposing $m_i=(0,0)$ gives producer $i$ a utility of $0$ which is his best response.

\textbf{Proof of Theorem \ref{feasibility_inelastic}}

Consider two cases.

\textit{Case 1:} For all $i \in I$, $p_i^* >0$.

Here, $p_i^* \geq 0$ is not binding. Therefore, at NE, for all $i \in I$,
\begin{eqnarray}
\label{NE_der_p}
\left.\frac{\partial u_i}{\partial p_i} \right|_{\textbf{m}^*} = \left.\frac{\partial t_i}{\partial p_i}\right|_{\textbf{m}^*} =\nonumber\\
-2(p^*_i-p^*_{i+1})-2{\zeta^*}^2 = 0.
\end{eqnarray} 
Summing up Eq. (\ref{NE_der_p}) over all $i$ we get:
\begin{eqnarray}
\label{Sum_der_p}
&&\sum _{i\in I} \left.\frac{\partial t_i}{\partial p_i}\right|_{\textbf{m}^*} \nonumber\\
&&=\sum_{i\in I} [ -2(p^*_i-p^*_{i+1}) - 2{\zeta^*}^2]\nonumber\\
&&= -2N{\zeta^*}^2 = 0.
\end{eqnarray}
But the last equation is achieved if and only if $\zeta^*=0$, i.e. if and only if $\sum_i e^*_i = D_0$.

\textit{Case 2:} $\exists i \in I$ $s.t.$ $p_i^*=0$.

We prove that this case can not be a non-trivial NE.

Since $p_i \geq 0$ is binding here, at NE we must have
\begin{equation}
\label{tax_p0_pos}
\left.\frac{\partial u_i}{\partial p_i}\right|_{\textbf{m}^*} \leq 0.
\end{equation}
Furthermore, using $p_i^* = 0$ in Eq. (\ref{NE_der_p}) we obtain
\begin{eqnarray}
\label{tax_der_p0}
\left.\frac{\partial u_i}{\partial p_i}\right|_{\textbf{m}^*} = 2 [p_{i+1}^* -{\zeta^*}^2]\leq 0.
\end{eqnarray}
Now assume $p_{i+1}^* > 0$. Then, from (\ref{NE_der_p}),
\begin{equation}
\label{p_i_1_not_binding_elastic}
\left.\frac{\partial u_{i+1}}{\partial p_{i+1}}\right|_{\textbf{m}^*} = 2(p_{i+2}^*-p_{i+1}^*)-2{\zeta^*}^2=0
\end{equation}
From  (\ref{p_i_1_not_binding_elastic}) it follows that
\begin{eqnarray}
p_{i+2}^* = p^*_{i+1}+{\zeta^*}^2 \geq p^{*}_{i+1} > 0.
\end{eqnarray}
Following the same argument, we obtain
\begin{eqnarray}
p_{i}^* = p^*_{i+1}+ N{\zeta^*}^2>0.
\end{eqnarray}
This contradicts the assumption of $p_i^*=0$. As a result, we should have $p_{i+1}^*=0$. Repeating the above argument we obtain $p_j^*=0, \quad \forall j \in I$. 

Next, if $p_i^* = p_{i+1}^*=0$, then from the utility function of producer $i$, it is clear that his best production amount is $e_i^*=0$. In the same way, $e_j^*=0, \forall j\in I$. But the bundle of zero production and zero price proposal for all producers is in contradiction with the assumption of non-trivial NE.

\textbf{Proof of Lemma \ref{lemma1_inelastic}}
First note that from proof of Theorem \ref{feasibility_inelastic}, for any non-trivial NE, we have $p_i^*>0, \forall i\in I$.
Using Eq. (\ref{zetaplus0}) in Eq. (\ref{NE_der_p}) we obtain
\begin{equation}
\left.\frac{\partial t_i}{\partial p_i}\right|_{\textbf{m}^*} = - 2(p^*_i-p^*_{i+1})= 0.
\end{equation}
Therefore,  
\begin{equation}
\label{price_same_2}
p^*_i=p^*_{i+1}=p^* \qquad \forall i\in I.
\end{equation}

Next, from (\ref{feasibility_1}), we get  $[\sum_{i\in I} e^*_i-D_0] = 0$, therefore,
\begin{equation}
\label{varesp0_2}
p^* [\sum_{i\in I} e^*_i-D_0] =0 \quad \forall i\in I.
\end{equation}
Finally, Eqs. (\ref{tax_form}), (\ref{zetaplus0}) and (\ref{price_same_2}) imply
\begin{equation}
t_i^* = p^* e_i^*,
\end{equation}
and
\begin{equation}
\left.\frac{\partial t_i}{\partial e_i}\right|_{\textbf{m}^*} = p^* - 2p^* {\zeta^*} \frac{\partial \zeta^*}{\partial e_i}= p^*.
\end{equation}

\textbf{Proof of Theorem \ref{Nash_Implementation_inelastic}}
Let $\textbf{m}^*=(\hat{\textbf{e}}^*, \textbf{p}^*)=(\hat{e}_1^*,\hat{e}_2^*,...,\hat{e}_N^*, p_1^*, p_2^*, ..., p_N^*)$  be a NE of the game induced by the mechanism. Then,  $\textbf{m}^*$ is a solution to every producer's profit maximization problem, that is,
\begin{eqnarray}
\label{Producer_Profit}
(\hat{e}_i^*, p_i^*)=\arg\max_{\hat{e}_i, p_i} & &- C_i(\hat{e}_i)+t_i\nonumber\\
\label{pro_capacity_constraint}
s.t & & 0 \leq \hat{e}_i \leq x_i \\
& & p_i \geq 0.
\end{eqnarray}
Call this problem \textbf{\textit{MAX2}}. The Lagrangian for this problem is
\begin{equation}
\label{Lagrangian_ind}
\mathcal{L} _{MAX2}= - C_i(\hat{e}_i)+t_i+\hat{\mu}_i (x_i-\hat{e}_i)+\hat{\nu}_i \hat{e}_i+\hat{\theta}_i p_i \\
\end{equation}
and the KKT conditions are, $\forall i \in I$,
\begin{eqnarray}
\label{KKT_pro_first}
-\left.\frac{\partial C_i}{\partial \hat{e}_i}\right|_{\textbf{m}^*}+\left.\frac{\partial t_i}{\partial \hat{e}_i}\right|_{\textbf{m}^*}-\hat{\mu}_i^*+\hat{\nu}_i^* &=& 0,\\
\frac{\partial t_i}{\partial p_i} |_{\textbf{m}^*} + \hat{\theta}_i^* &=& 0,\\
\hat{\mu}^*_i(x_i-\hat{e}_i^*)&=&0,\\
\hat{\nu}^*_i \hat{e}_i^* &=&0,\\
\hat{\theta}^*_i p_i^*&=&0,\\
\hat{\nu}_i^* &\geq& 0,\\
\label{pro_KKT_mu}
\hat{\mu}_i^* &\geq&0,\\
\label{KKT_pro_last}
\hat{\theta}_i^* &\geq& 0.
\end{eqnarray}

To show that the allocation $e_i^*=\hat{e}_i^*$, $i\in I$, corresponding to $m^*$ is a solution of the centralized problem \textit{\textbf{MAX2}}, we construct the KKT parameters of the centralized problem based on the producers' profit maximization KKT parameters as follows.
\begin{eqnarray}
\label{KKT_equivalency_1}
\lambda^*&=&p^*,\\
\mu_i^*&=&\hat{\mu}_i^*\quad \forall i\in I,\\
\nu_i^*&=&\hat{\nu}_i^*\quad \forall i\in I .
\end{eqnarray}
Then, Eqs. (\ref{KKT_pro_first})-(\ref{KKT_pro_last}) along with lemma \ref{lemma1_inelastic} show that Eqs. (\ref{KKT_MAX1_first})-(\ref{KKT_MAX1_last}) are satisfied. The assertion of Theorem (\ref{Nash_Implementation_inelastic}) follows, since $\textbf{m}^*$ is an arbitrary non-trivial NE of the game induced by the mechanism.

\textbf{Proof of Theorem \ref{Strong_Implementation_inelastic}}
Consider the KKT conditions for problem \textbf{\textit{MAX2}} expressed by Eqs. (\ref{KKT_MAX1_first})-(\ref{KKT_MAX1_last}) and set, $\quad \forall i\in I$,
\begin{eqnarray}
\label{KKT_equivalency_2}
p_i^*&=&\lambda^*,\\
\hat{\mu}_i^*&=&\mu_i^*,\\
\hat{\nu}_i^*&=&{\nu}_i^*,\\
\hat{\theta}_i^*&=&0.
\end{eqnarray}
Eqs. (\ref{KKT_MAX1_first})-(\ref{KKT_MAX1_last}) show that Eqs. (\ref{KKT_pro_first})-(\ref{KKT_pro_last}) are satisfied under the above selection.

\textbf{Proof of Theorem \ref{Individual_Rationality_inelastic}}

Consider 3 cases.

\textit{Case 1:} $e_i^*=0$.

Then Eq. (\ref{tax_equ}) results in
\begin{equation}
u_i (\textbf{e}^*, \textbf{p}^*) = -C_i(e_i^*)-t_i^*= -C_i(0)-p_i^*\times 0 = 0.
\end{equation}
Therefore, the NE outcome is weakly preferred to the initial allocation.

\textit{Case 2:} $ 0 < e_i^* < x_i $. 

The constraint in Eq. (\ref{pro_capacity_constraint}) is not binding and therefore the corresponding $\hat{\mu}_i^*$ and $\hat{\nu}_i^*$ are $0$. Then, Eqs. (\ref{tax_equ}), (\ref{tax_der_equ}) and (\ref{KKT_pro_first}) along with $\hat{\mu}_i^*=\hat{\nu}_i^*=0$ result in 
\begin{equation}
\label{Utility at equ 2_inelastic}
u_i(\textbf{e}^*, \textbf{p}^*) =  -C_i(e_i^*)+t_i^* = -C_i(e_i^*) + C_i^{'}(e_i^*) e_i^*.
\end{equation}
Furthermore, from (\ref{C_0})-(\ref{C_second_der}), for the convex and increasing function $C_i$
\begin{equation}
\label{C_less_inelastic}
C_i(e_i) < C_i^{'}(e_i) e_i \qquad \text{for any}\quad e_i>0.
\end{equation}
Combining (\ref{Utility at equ 2_inelastic}) and (\ref{C_less_inelastic}) we get, 
\begin{equation}
u_i(\textbf{e}^*, \textbf{t}^*) = -C_i(e_i^*) + C_i^{'}(e_i^*) e_i^* >0.
\end{equation}

\textit{Case 3:} $ e_i^* = x_i $. 

Since the constraint $e_i \leq x_i$ is binding, $\hat{\mu}_i^*>0$, $\hat{\nu}_i^*=0$, and Eqs. (\ref{C_0}), (\ref{C_first_der}), (\ref{tax_der_equ}) and (\ref{KKT_pro_first}) imply
\begin{equation}
\label{price_upper_const_inelastic}
p^*= \frac{\partial t_i}{\partial e_i} = C_i^{'}(e_i^*)+\hat{\mu}^* > C_i^{'}(e_i^*).
\end{equation}
Inequality (\ref{price_upper_const_inelastic}) along with (\ref{tax_equ}) and (\ref{C_less_inelastic}) result in
\begin{eqnarray}
&&u_i(\textbf{e}^*, \textbf{t}^*)   -C_i(e_i^*)+t_i^*=\nonumber\\
&&-C_{i}(e_i^*)+p^* e_i^* > -C_i(e_i^*) + C_i^{'}(e_i^*) e_i^* > 0.
\end{eqnarray}

\textbf{Proof of Theorem \ref{budget_balance_inelastic}}

Producer $i$ receives $t_i$ and demand pays $\sum_{i\in I }t_i$. Therefore, the sum of all payments adds up to zero at every message proposal.

\textbf{Proof of Theorem \ref{budget_efficiency_inelastic}}

Consider producer $i \in I$ for which the production capacity constraints are not binding, i.e. $0 < e_i^* < x_i$; therefore, $\hat{\mu}_i^*=\hat{\nu}_i^*=0$ in Eq. (\ref{KKT_pro_first}). This along with Eq. (\ref{tax_der_equ}) imply that
\begin{equation}
\label{price_marginal_cost_inelastic}
p^* = p_i^* = u_d^{'}(\sum_{i\in I} e_i^*).
\end{equation}
Then, because of Eq. (\ref{tax_equ})
\begin{equation}
\sum_i t_i^* = p^*\times \sum_i e_i^*,
\end{equation}
which along with (\ref{price_marginal_cost_inelastic}) means that the demand side pays marginal cost of production multiplied by quantity.

\section{\textbf{Example for Electricity Pooling Market with Inelastic Demand}}
\label{appendix_C}

Here, we provide an example of electricity pooling market with inelastic demand.

As we proposed in Section \ref{Conclusion}, we currently don't have an algorithm for computing the NE of the game induced by the mechanism. Nevertheless, we have proven that the production profiles corresponding to all non-trivial NE of the game induced by the mechanism are optimal solutions of the corresponding centralized problems. Thus, we obtain these production profiles as the solution of the centralized problem.

\begin{example}
\label{example_1}

Consider a network of four producers with following cost functions
\begin{eqnarray}
C_1(e_1)&=&2e_1+e_1^2\\
C_2(e_2)&=&3e_2+e_2^3\\
C_3(e_3)&=&4e_3+e_3^4\\
C_4(e_4)&=&5e_4+e_4^2
\end{eqnarray}
and the following capacities,
\begin{equation}
x_1=x_2=x_3=x_4=2.
\end{equation}
The demand is inelastic with $D_0=4MWh$.

The social welfare function, Eq. (\ref{social_welfare_inelastic}), will be
\begin{eqnarray} 
W(e_1, e_2, ..., e_4, D_0)= u_d(D_0) - \sum_{i=1,2,...,4} C_i(e_i) =\nonumber\\
u_d(D_0) - [2e_1+e_1^2+3e_2+e_2^3+4e_3+e_3^4+5e_4+e_4^2].
\end{eqnarray}
and the corresponding problem \textbf{\textit{MAX1}} is 
\begin{align*}
\max_{e_i , i\in I}& - [2e_1+e_1^2+3e_2+e_2^3+4e_3+e_3^4+5e_4+e_4^2]\nonumber\\
&s.t. 0 \leq e_i \leq 2 \nonumber\\
& \sum_{i\in I} e_i \geq 4.
\end{align*}
Solving the above maximization, the optimal procurement is as follows.
\begin{eqnarray}
e^*_1&=&2\\
e^*_2&=& 1.12\\
e_3^*&=&0.88\\
e_4^*&=&0
\end{eqnarray}

The corresponding game form induced by the mechanism designed for inelastic demand has the following form:

\underline{Message space}
\begin{equation}
\mathcal{M}_i:= [0, 2]\times \mathbb{R}_{+}, i\in \{1,2,3,4\}
\end{equation}
and $m_i\in \mathcal{M}_i$ is of the form 
\begin{equation}
m_i=(\hat{e}_i, p_i)
\end{equation}
where $\hat{e}_i$ denotes the amount of energy producer $i$ proposes to produce, and $p_i$ denotes the price producer $i$ demands per unit of energy.

\underline{Allocation Space}  

Let $\mathcal{A}$ be
\begin{equation}
\mathcal{A}:=(\mathcal{A}_1 \otimes \mathcal{A}_2 \otimes ... \otimes \mathcal{A}_N), 
\end{equation}
where $\mathcal{A}_i$ is producer $i$'s allocation space
\begin{equation}
\mathcal{A}_i:= [0, 2]\times \mathbb{R}, i\in I,
\end{equation}
and $a_i\in \mathcal{A}_i$ is of the form 
\begin{equation}
a_i=(e_i, t_i),
\end{equation}
where ${e}_i$ denotes the amount of energy producer $i$ is scheduled to produce, and $t_i$ denotes the subsidy (respectively tax) producer $i$ should receive (respectively pay).

\underline{Outcome function} $h: \mathcal{M} \rightarrow \mathcal{A}$

For each $m \in \mathcal{M}$ we have 
\begin{equation}
h(m)=({\textbf{e}}, \textbf{t})=({e}_1,...,{e}_N,{t}_1,...,{t}_N)
\end{equation}
where
\begin{eqnarray}
{e_i} &=& \hat{e_i} \\
t_i &=& p_{i+1} e_i - (p_i-p_{i+1})^2 - 2 p_i \zeta^2\\
\zeta &=&|0, D_0-\sum_{i\in I} e_i|\nonumber\\
&=&|4-\sum_{i\in I} e_i|\\
p_{5}&:=&p_{1}.
\end{eqnarray}
We proceed to interpret the mechanism and to analyze its properties.

This game has a NE of the form $\hat{e}^*_i=e_i^*, p_i^*=p^*$. Note that here the first producer capacity limit is binding. The producer 4 also has a binding production constraint but from below. The price will be the marginal cost of production, which is the marginal cost of production of the second and the third producers. The price and the tax payments at equilibrium will be the following.
\begin{eqnarray}
p^*=\left. \frac{\partial C_2}{\partial e_2}\right|_{e^*}=\left. \frac{\partial C_3}{\partial e_3}\right|_{e^*}&=&6.5 \$/MWh \\
t_1^*&=&13 \$ \\
t_2^*&=&7.28 \$ \\
t_3^*&=&5.72 \$ \\
t_4^*&=&0 \$ 
\end{eqnarray}

Form Eqs. (\ref{utility_consumers}), the utility of producers at equilibrium is the following.
\begin{eqnarray}
u_1&=&5 \$\\
u_2&=&2.53 \$\\
u_3&=&1.59 \$\\
u_4&=&0 \$
\end{eqnarray}
The outcome of the game form is individually rational, budget balanced and price efficient. 
\end{example}
\end{appendices}

\bibliography{IEEEabrv,mybib,collection}
\end{document}